# Area Efficient Hardware Implementation of Elliptic Curve Cryptography by Iteratively Applying Karatsuba's Method

Zoya Dyka and Peter Langendoerfer
IHP, Im Technologiepark 25, 15236 Frankfurt (Oder), Germany
langendoerfer@ihp-microelectronics.com

## Abstract

*Securing communication channels is especially needed in wireless environments. But applying cipher mechanisms in software is limited by the calculation and energy resources of the mobile devices. If hardware is applied to realize cryptographic operations cost becomes an issue. In this paper we describe an approach which tackles all these three points. We implemented a hardware accelerator for polynomial multiplication in extended Galois fields (GF) applying Karatsuba's method iteratively. With this approach the area consumption is reduced to 2.1 mm$^2$ in comparison to. 6.2 mm$^2$ for the standard application of Karatsuba's method i.e. for recursive application. Our approach also reduces the energy consumption to 60 per cent of the original approach. The price we have to pay for these achievement is the increased execution time. In our implementation a polynomial multiplication takes 3 clock cycles whereas the recurisve Karatsuba approach needs only one clock cycle. But considering area, energy and calculation speed we are convinced that the benefits of our approach outweigh its drawback.*

## 1. Introduction

**Motivation** Mobile devices are penetrating our every day life. More and more sensitive information is exchanged between mobile nodes and between mobile and fixed communication endpoints. This data exchange is normally protected by cipher mechanisms. But due to the scarce resources of mobile nodes, exhaustive use of cryptographic means is infeasible. This holds especially true for public key cryptography, which is normally used to establish a secure channel between the communicating parties as well as for providing digital signatures. Hardware accelerators for public key cryptography operations are ideal means to reduce the calculation time as well as the energy consumption. But, a straight forward realization of cryptographic operations results in a relatively large area consumption, which makes the application of hardware accelerators economically infeasible. Thus, our design constraints were:

- Calculation time,
- Energy consumption, and
- Area consumption.

We decided to use Elliptic Curve Cryptography (ECC) since it guarantees the same security level as RSA does but with significant shorter keys. In addition to this the ECC operations are faster than those of RSA [1]. We selected B-233 over Galois field GF($2^{233}$) which is recommended by NIST [2] and well suited to be implemented in hardware.

Despite ECC is less computational intensive than RSA it still requires a significant effort in terms of energy and time. In this paper we concentrate on the area efficient realization of basic mathematical operations, which are used in ECC.

**ECC Background** In order to calculate the product of two 233 bit long operands, denoted 'kP'. Here $P$ is a point on an elliptic curve (EC) and $k$ is a large number. The 'kP' multiplication is based on point doubling and point addition. All these EC point operations are based on addition, subtraction, squaring, multiplication and division in a chosen GF. The basic operations in GF($2^{233}$) are addition, squaring, multiplication and division of polynomials. Addition of polynomials is equivalent to a bit-wise XOR operation. Squaring and multiplication require two steps: squaring/multiplication itself and reduction of the result. Reduction is done using so-called irreducible polynomials and it is a fast operation in GF($2^n$). The irreducible polynomial for B-233 is the trinomial: $f(x) = x^{233} \oplus x^{74} \oplus 1$ [1]. Division of polynomials usually is done in two steps: first identifying the inverse of the divisor using the irreducible polynomial, and second multiplying the inverse with the dividend. Multiplication and division of polynomials require the major part of the calculation time.
In this paper we are concentrating on polynomial multiplication, since our long term goal is to implement a Montgomery multiplier for the 'kP' operation. The Montgomery method requires only one polynomial division for 'kP', so that the major effort comes from the multiplication.

**Contribution and structure of this paper** In this paper we show that an iterative application of the Karatsuba method provides very good results with respect to the following three parameters: calculation time, area consumption and energy consumption. With our iterative hardware solution, the chip area needed to calculate the product of two 233 bit long operands, is 2.1 $mm^2$ whereas the standard application of Karatsuba's method needs 6.2 $mm^2$. Our approach also reduces the energy consumption to 60 per cent of the original approach. The price we have to pay for these achievements is the increased execution time. In our implementation a polynomial multiplication takes 3 clock cycles whereas the original one needs only one clock cycle.

The rest of this paper is structured as follows. Section 2 contains a short description of implemented methods. We propose to use the Karatsuba's formula for polynomial multiplication iteratively. The detailed description of our approach is given in Section 3. Section 4 discusses the hardware realization of our approach and provides measurement results. We conclude the paper with a short discussion of our results and an outlook on further research steps.

## 2. State of the art

In this section we describe methods for polynomial multiplication in polynomial basis. We implemented these methods and different combinations of them to realize our own approach and to benchmark our solution.

[1] Note: In GF($2^n$) addition and subtraction are XOR operations. Due to this and for simpler understanding of formulas we change the usual representation of polynomials $A(x) = \sum_{i=0}^{n-1} a_i x^i$ to $A(x) = \bigoplus_{i=0}^{n-1} a_i x^i$. In rest of this paper we denote XOR operation as '⊕'. The symbol '+' means always an ordinary addition.

### 2.1. Polynomial multiplication

The product of two polynomials

$$A(x) = \bigoplus_{i=0}^{n-1} a_i x^i \quad \text{and} \quad B(x) = \bigoplus_{i=0}^{n-1} b_i x^i$$

is the polynomial: $C(x) = A(x) \cdot B(x) = \bigoplus_{i=0}^{2n-2} c_i x^i$,

where $c_i = \bigoplus_{k+l=i} a_k \cdot b_l$, i.e.:

$$\begin{aligned}
c_0 &= a_0 \cdot b_0 \\
c_1 &= a_1 \cdot b_0 \oplus a_0 \cdot b_1 \\
&\vdots \\
c_{n-1} &= a_{n-1} \cdot b_0 \oplus a_{n-2} \cdot b_1 \oplus \ldots \oplus a_0 \cdot b_{n-1} \\
&\vdots \\
c_{2n-3} &= a_{n-1} \cdot b_{n-2} \oplus a_{n-2} \cdot b_{n-1} \\
c_{2n-2} &= a_{n-1} \cdot b_{n-1}
\end{aligned} \tag{1}$$

The straight forward implementation of formula (1) requires $n^2$ partial multiplication and $(n-1)^2$ XOR operations of partial products in order to calculate $c_i$. All operands in formula (1) are only *one-bit* long. In case of using EC B-233 both polynomials *A(x)* and *B(x)* are *233-bit* long. It means that in total $233^2$ *one-bit* partial multiplications and $232^2$ XOR operations are required.

### 2.2. Karatsuba based methods

**Original Karatsuba's method** For polynomial multiplication with original Karatsuba method [3] both operands have to be fragmentized into two equal parts. If the length $n$ of operands is odd, they have to be padded with leading '0'. So, operands can be written as[2]:

$$\begin{aligned}
A(x) &= a_{n-1} \ldots a_{\frac{n}{2}} a_{\frac{n}{2}-1} \ldots a_1 a_0 = a_{n-1} \ldots a_{\frac{n}{2}} \cdot x^{\frac{n}{2}} \oplus a_{\frac{n}{2}-1} \ldots a_1 a_0 = \\
&= a^1 \cdot x^{\frac{n}{2}} \oplus a^0
\end{aligned} \tag{2}$$

[2] We denote as $\boldsymbol{a_i}$ the $i^{th}$ bit and as $\boldsymbol{a^i}$ the $i^{th}$ segment of operand *A(x)*.

The polynomial B(x) is represented in the same way. The Karatsuba's formula for the product $C(x)=A(x)\cdot B(x)$ is

$$C(x) = a^0b^0 \oplus \left[a^0b^0 \oplus a^1b^1 \oplus (a^0 \oplus a^1)(b^0 \oplus b^1)\right] \cdot x^{\frac{n}{2}} \oplus \oplus a^1b^1 \cdot x^n \tag{3}$$

In order to calculate the partial products $a^ib^i$ Karatsuba's formula can be applied recursively. In this case it requires in total $s^{\log_2 3} = s^{1.58}$ partial multiplications, where $s$ is the number of segments. This method can be used to speed up software as well as hardware implementations. Usually in software implementations the Karatsuba's approach is applied until both operands have a size of one word.

In Bailey and Paar [4] a new scheme how to apply Karatsuba's idea was proposed. In this scheme the operands are divided into three parts. Throughout the rest of this paper we denote this method as ***Bailey's method***. It requires 6 partial multiplications of *n/3-bit* long operands. This method can be combined with the original Karatsuba formula for operands, whose length is divisible by six.

## 3. Iterative Application of the Karatsuba Approach

The major point in our approach is to apply the original version Karatsuba's method iteratively. We denote this as ***Iterative-Karatsuba method***. The major benefits of this approach are:

- a smaller area consumption of the hardware accelerators due to the fact that partial multiplications can be performed serially
- a reduced number of XOR operations compared with the recursive variant of Karatsuba's method.

We explain our idea of the iterative application of Karatsuba's formula using an example in which the operands are split up into four segments. First of all, we use the original Karatsuba formula to obtain the expression for a product, in which only 1-segment long operands for partial multiplication are used.

So, at the beginning we have two operands, each of them *4n-bit* long. We fragment each operand into two *2n-bit* long parts:

$$\begin{aligned} A(x) &= a^3a^2a^1a^0 = a^3a^2 \cdot x^{2n} \oplus a^1a^0 \\ B(x) &= b^3b^2b^1b^0 = b^3b^2 \cdot x^{2n} \oplus b^1b^0 \end{aligned} \tag{4}$$

The result of applying Karatsuba's formula is:

$$\begin{aligned} C(x) &= a^1a^0 \cdot b^1b^0 \oplus \\ &\oplus \left[a^1a^0 \cdot b^1b^0 \oplus a^3a^2 \cdot b^3b^2 \oplus a^{13}a^{02} \cdot b^{13}b^{02}\right] \cdot x^{2n} \oplus \\ &\oplus a^3a^2 \cdot b^3b^2 \cdot x^{4n} \end{aligned} \tag{5}$$

*where*

$$\begin{aligned} a^{13}a^{02} &= a^{13} \cdot x^n \oplus a^{02} = (a^1 \oplus a^3) \cdot x^n \oplus (a^0 \oplus a^2) = \\ &= (a^1 \cdot x^n \oplus a^0) \oplus (a^3 \cdot x^n \oplus a^2) = a^1a^0 \oplus a^3a^2 \end{aligned} \tag{6}$$

*and*

$$b^{13}b^{02} = b^1b^0 \oplus b^3b^2 \tag{7}$$

Every 2-segments element is: $a^ia^j = a^i \cdot x^n \oplus a^j$. So, for each partial multiplication from (6) and (7) we use the Karatsuba's formula again. The final result is given in formula (8).

$$\begin{aligned} C(x) = {} & a^3 \cdot b^3 \cdot x^{6n} \oplus (a^2 \cdot b^2 \oplus a^3 \cdot b^3 \oplus a^{23} \cdot b^{23}) \cdot x^{5n} \oplus \\ & \oplus (a^1 \cdot b^1 \oplus a^2 \cdot b^2 \oplus a^3 \cdot b^3 \oplus a^{13} \cdot b^{13}) \cdot x^{4n} \oplus \\ & \oplus (a^0 \cdot b^0 \oplus a^1 \cdot b^1 \oplus a^2 \cdot b^2 \oplus a^3 \cdot b^3 \oplus \\ & \quad \oplus a^{01} \cdot b^{01} \oplus a^{02} \cdot b^{02} \oplus a^{13} \cdot b^{13} \oplus a^{23} \cdot b^{23} \oplus \\ & \quad \oplus a^{0123} \cdot b^{0123}) \cdot x^{3n} \oplus \\ & \oplus (a^0 \cdot b^0 \oplus a^1 \cdot b^1 \oplus a^2 \cdot b^2 \oplus a^{02} \cdot b^{02}) \cdot x^{2n} \oplus \\ & \oplus (a^0 \cdot b^0 \oplus a^1 \cdot b^1 \oplus a^{01} \cdot b^{01}) \cdot x^n \oplus a^0 \cdot b^0 \end{aligned} \tag{8}$$

Each of the operands is 1-segment long, so that the resulting partial product is *(2n-1)-bit* long. We denote the bits from *n-1* to *0* of the product $a^i \cdot b^i$ as $a^ib^i[0]$ and the bits from *2n-1* to *n* as $a^ib^i[1]$:

$$a^i \cdot b^i = a^ib^i[1] \cdot x^n \oplus a^ib^i[0] \tag{9}$$

Using the notation introduced in (9) we can represent formula (8) as given in table 1.

**Table1. Representation of formula (8)**

| partial products | segments of result | | | | | | | |
|---|---|---|---|---|---|---|---|---|
| $a^0 \cdot b^0$ [0] | | | | | ⊕ | ⊕ | ⊕ | ⊕ |
| $a^0 \cdot b^0$ [1] | | | | ⊕ | ⊕ | ⊕ | ⊕ | |
| $a^1 \cdot b^1$ [0] | | | | ⊕ | ⊕ | ⊕ | ⊕ | |
| $a^1 \cdot b^1$ [1] | | | ⊕ | ⊕ | ⊕ | ⊕ | | |
| $a^2 \cdot b^2$ [0] | | | ⊕ | ⊕ | ⊕ | ⊕ | | |
| $a^2 \cdot b^2$ [1] | | ⊕ | ⊕ | ⊕ | ⊕ | | | |
| $a^3 \cdot b^3$ [0] | | ⊕ | ⊕ | ⊕ | ⊕ | | | |
| $a^3 \cdot b^3$ [1] | ⊕ | ⊕ | ⊕ | ⊕ | | | | |
| $(a^0 \oplus a^1)\cdot(b^0 \oplus b^1)$ [0] | | | | | ⊕ | | ⊕ | |
| $(a^0 \oplus a^1)\cdot(b^0 \oplus b^1)$ [1] | | | | ⊕ | | ⊕ | | |
| $(a^0 \oplus a^2)\cdot(b^0 \oplus b^2)$ [0] | | | | | ⊕ | ⊕ | | |
| $(a^0 \oplus a^2)\cdot(b^0 \oplus b^2)$ [1] | | | | ⊕ | ⊕ | | | |
| $(a^1 \oplus a^3)\cdot(b^1 \oplus b^3)$ [0] | | | | ⊕ | ⊕ | | | |
| $(a^1 \oplus a^3)\cdot(b^1 \oplus b^3)$ [1] | | | ⊕ | ⊕ | | | | |
| $(a^2 \oplus a^3)\cdot(b^2 \oplus b^3)$ [0] | | | ⊕ | | ⊕ | | | |
| $(a^2 \oplus a^3)\cdot(b^2 \oplus b^3)$ [1] | | ⊕ | | ⊕ | | | | |
| $(a^0 \oplus a^1 \oplus a^2 \oplus a^3)\cdot(b^0 \oplus b^1 \oplus b^2 \oplus b^3)$ [0] | | | | | ⊕ | | | |
| $(a^0 \oplus a^1 \oplus a^2 \oplus a^3)\cdot(b^0 \oplus b^1 \oplus b^2 \oplus b^3)$ [1] | | | | ⊕ | | | | |
| $C(x) =$ | $c^7$ | $c^6$ | $c^5$ | $c^4$ | $c^3$ | $c^2$ | $c^1$ | $c^0$ |

All columns in table 1 which are nested under the topic '*segments of result*' in represent a certain segment $c^i$. For each partial product two lines are given in Table 1, one line representing the lower ($a^x b^x$[0]), and a second one representing the upper part ($a^x b^x$[1]) of the product as specified above. The segment $c^i$ can be calculated by XOR-ing all lines in the table 1, which contain the symbol '⊕' in the column of $c^i$. For example $c^5$ can be calculated as follows:

$$c^5 = a^1b^1[1] \oplus a^2b^2[0] \oplus a^2b^2[1] \oplus a^3b^3[0] \oplus a^3b^3[1] \oplus ((a^1 \oplus a^3)(b^1 \oplus b^3)[1]) \oplus ((a^2 \oplus a^3)(b^2 \oplus b^3)[0]) \quad (10)$$

Each segment $c^i$ can be calculated iteratively i.e. step by step as we calculate the partial products starting with $a^0b^0$ down to $(a^0 \oplus a^1 \oplus a^2 \oplus a^3)\cdot(b^0 \oplus b^1 \oplus b^2 \oplus b^3)$. We then start to calculate the segments of products using the already received results. For example:

| Step 1 | Step 2 | … And so on to Step 9 |
|---|---|---|
| $c^0 = a^0b^0[0]$ | $c^1 = c^1 \oplus a^1b^1[0]$ | |
| $c^1 = a^0b^0[0] \oplus a^0b^0[1]$ | $c^2 = c^2 \oplus a^1b^1[0] \oplus a^1b^1[1]$ | |
| $c^2 = a^0b^0[0] \oplus a^0b^0[1]$ | $c^3 = c^3 \oplus a^1b^1[0] \oplus a^1b^1[1]$ | |
| $c^3 = a^0b^0[0] \oplus a^0b^0[1]$ | $c^4 = c^4 \oplus a^1b^1[0] \oplus a^1b^1[1]$ | |
| $c^4 = a^0b^0[1]$ | $c^5 = a^1b^1[1]$ | |

This iterative calculation of the $C(x)$ reduces the area of our hardware multiplier. We need only one partial multiplier for 1-segment long operands. After each new clock this multiplier delivers the next partial product. In that way the segments of product $C(x)$ are collected. For the above given example this means after 9 clock cycles all segments contain the correct product of the polynomial multiplication.

Additionally we exploit another 'iterative possibility': we do not need to calculate all segments of $C(x)$ separately. We can use $c^0$ to determine $c^1$ after the first clock, $c^1$ for $c^2$ after second clock, and so on (see table 2). This iterative calculation reduces the number of XOR operations to 29 compared to 42 XOR operations if the calculation of every $c^i$ is done separately.

In a similar way we applied our iterative approach to Bailey's method, which we call Iterative-Bailey throughout the rest of this paper.

**Table 2. Exact operation sequence of our hardware implementation of formula (8)**

| clock | obtained partial product | sequence of operations |
|---|---|---|
| 1 | $pr = a^0 \cdot b^0$ | $c^0 = pr[0]$<br>$c^1 = pr[1]$ |
| 2 | $pr = a^1 \cdot b^1$ | $c^1 = c^1 \oplus c^0 \oplus pr[0]$<br>$c^2 = pr[1]$ |
| 3 | $pr = a^2 \cdot b^2$ | $c^2 = c^2 \oplus c^1 \oplus pr[0]$<br>$c^3 = pr[1]$ |
| 4 | $pr = a^3 \cdot b^3$ | $c^3 = c^3 \oplus c^2 \oplus pr[0] \oplus pr[1]$<br>$c^7 = pr[1]$ |
| 5 | $pr = (a^0 \oplus a^1)\cdot(b^0 \oplus b^1)$ | $c^6 = c^3 \oplus c^2$<br>$c^5 = c^3 \oplus c^1$<br>$c^4 = c^3 \oplus c^0 \oplus pr[1]$<br>$c^3 = c^3 \oplus c^7 \oplus pr[0]$<br>$c^2 = c^2 \oplus pr[1]$<br>$c^1 = c^1 \oplus pr[0]$ |
| 6 | $pr = (a^0 \oplus a^2)\cdot(b^0 \oplus b^2)$ | $c^3 = c^3 \oplus pr[0] \oplus pr[1]$<br>$c^2 = c^2 \oplus pr[0]$<br>$c^4 = c^4 \oplus pr[1]$ |
| 7 | $pr = (a^1 \oplus a^3)\cdot(b^1 \oplus b^3)$ | $c^4 = c^4 \oplus pr[0] \oplus pr[1]$<br>$c^3 = c^3 \oplus pr[0]$<br>$c^5 = c^5 \oplus pr[1]$ |
| 8 | $pr = (a^2 \oplus a^3)\cdot(b^2 \oplus b^3)$ | $c^3 = c^3 \oplus pr[0]$<br>$c^5 = c^5 \oplus pr[0]$<br>$c^4 = c^4 \oplus pr[1]$<br>$c^6 = c^6 \oplus pr[1]$ |
| 9 | $pr = (a^0 \oplus a^1 \oplus a^2 \oplus a^3)\cdot$<br>$\cdot(b^0 \oplus b^1 \oplus b^2 \oplus b^3)$ | $c^3 = c^3 \oplus pr[0]$<br>$c^4 = c^4 \oplus pr[1]$ |

## 4. Hardware implementation

In this section we will present the design and the key parameters of our hardware realization of the Iterative Karatsuba approach.

The design of the Iterative Karatsuba accelerator consists of three major parts (see Fig. 1):

- S*election* block feeds certain parts of both operands into the *Partial Multiplier*, for each new clock signal.
- *Partial Multiplier* block calculates the partial product of the operands delivered by the selection block and provides the results to the *product accumulation block*.
- *Product Accumulation* block computes the final product from the partial products it receives from the partial multiplier. The theoretical basis and exact operation sequence is discussed in detail in Section 3.

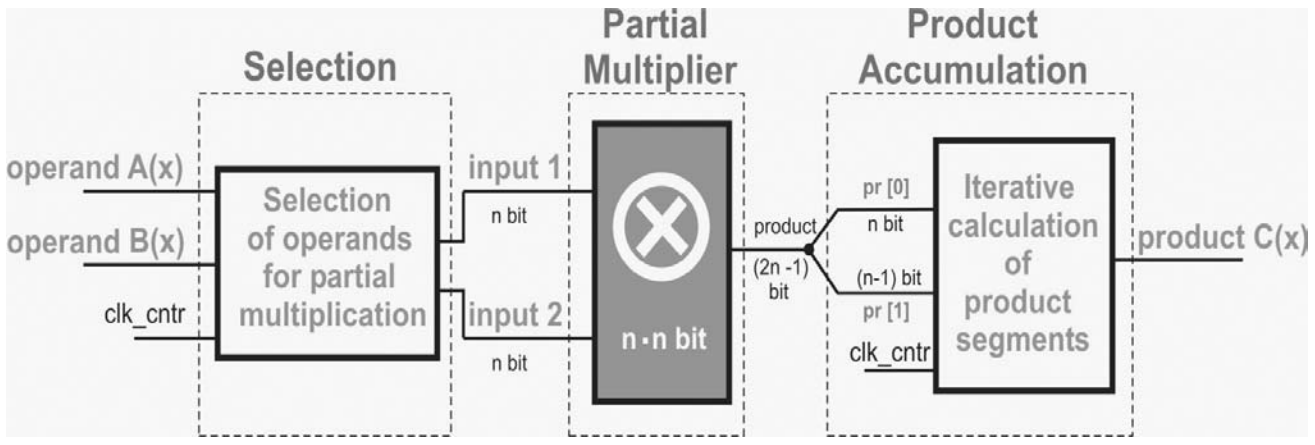

**Figure 1: Block diagram of our Iterative-Karatsuba multiplier**

The performance, chip area and energy consumption of a polynomial multiplier are dominated by the partial multiplier which is used. The larger the input signals of the partial multiplier may be, the faster the partial multiplier is. But this also results in a relatively large area consumption. So, the design decision to make seems to be straight forward: calculation time versus chip area. This is true as long as only the partial multiplier is considered. But for the polynomial multiplier also the area of the selection and the product accumulation block have to be taken into account. The chip area needed for the accumulation block depends on the area of the partial multiplier in an inverse proportional manner, i.e. the smaller the partial multiplier the larger the accumulation block. This results from the fact that in case of small partial multipliers more intermediary results have to be stored for the final calculation of the polynomial product. For example the size of the accumulation block is 0,649 $mm^2$ if the partial multiplier accepts 128 bit long operands, and 1,466 $mm^2$ if the maximum length of the operands is 32 bits.

In order to determine the most appropriate design for a polynomial multiplier we realized several partial multipliers. We realized 3 one-clock partial multipliers for our iterative Karatsuba as well as for our iterative Bailey approach. These partial multipliers accept operands with a maximal length of 128, 64 and 32 bits respectively. They were synthesized with a library of our in-house 0.25 µm CMOS-Technology [5]. Table 3 shows the area, the time and energy consumption for each of these six partial multipliers. These values stem from the Design analyzer tool from Synopsys [6].

**Table 3. Parameters of synthesized partial multipliers**

| Name of partial multiplier (PM) | Length of input values, bits | Area, $mm^2$ | time, ns | Energy/clock, pW·s |
|---|---|---|---|---|
| *k128_k64_k32_k16_sh8* | *128* | 1.620410 | 12.53 | 1394.4000 |
| *k64_k32_k16_sh8* | *64* | 0.514759 | 8.99 | 404.3913 |
| *k32_k16_sh8* | *32* | 0.159006 | 5.62 | 108.2011 |
| *p81_p27_sh9* | *81* | 0.896391 | 7.98 | 692.0033 |
| *p39_sh13* | *39* | 0.264672 | 5.97 | 179.4565 |
| *p27_sh9* | *27* | 0.133616 | 4.80 | 88.4779 |

In order to benchmark our approach we realized polynomial multipliers using the following approaches:

- Iterative Karatsuba
- Iterative Bailey
- Original Karatsuba (recursive)
- Original Bailey (recursive)

For the first two approaches, i.e. for our own iterative approaches, we realized three polynomial multipliers using different partial multipliers (see table 3) in order to see how the partial multiplier influences the overall parameters. We named these multipliers so that the name indicates the applied method. For example, the name *iterative_Karatsuba_8segments* means: Iterative-Karatsuba fragmentizing incoming operands into 8 segments.

In the two recursive multipliers the original Karatsuba and the Bailey formula are applied down to *one-bit* operands. Both multipliers deliver the polynomial product after one clock cycle. They differ in the length of the input operands. The Karatsuba multiplier expects always two 256 bit long input values whereas the Bailey multiplier expects two *243-bit* long input values.

Since we are going to use these multipliers for EC B-233 the two input values will be only *233-bit* long. Therefore the operands were padded with leading 0's if it was necessary. The result of the multiplication is always *465-bit* long.

We synthesized all polynomial multipliers using a library of our in-house 0.25µm CMOS-Technology [5]. We obtained the data represented in these tables with different kinds of reports from the Synopsys "Design Analyzer" [6]. The parameters of the implemented polynomial multipliers are given in Table 4. Our results clearly indicate that an iterative application of the original Karatsuba and Bailey approach significantly reduces the chip area. If the number of iterations is kept small, our approach also helps to reduce the energy consumption. In those designs the decision is less area and less energy versus slower execution time. Increasing the number of iterations helps to reduce the chip area needed, but it also leads to an increased power consumption and an increased calculation time. So, these implementations are beneficial only if cost is the dominating parameter.

**Table 4. Parameters of synthesized polynomial multipliers**

| Name of multiplier | Area, S, $mm^2$ | Number of clocks, N | Period, T, ns | Power, P, mW | Energy, E=T·N·P, pW·s |
|---|---|---|---|---|---|
| *iterative_Karatsuba_ 2segments (PM - k128_k64_ k32_k16_sh8)* | 2.18 | 3 | 15 | 98.89 | 4450.1 |
| *iterative_Karatsuba_ 4segments (PM - _k64_k32_k16_sh8)* | 1.52 | 9 | 10 | 105.48 | 9493.2 |
| *iterative_Karatsuba_ 8segments (PM -_k32_k16_sh8)* | 1.67 | 27 | 9 | 107.63 | 26154.1 |
| *iterative_Bailey_ 3segments (PM - p81_p27_sh9)* | 2.12 | 6 | 10 | 148.16 | 8889.6 |
| *iterative_Bailey_ 6segments (PM - p39_sh13)* | 1.60 | 18 | 9 | 110.46 | 17894.5 |
| *iterative_Bailey_ 9segments (PM - p27_sh9)* | 1.71 | 36 | 9 | 103.35 | 33485.4 |
| *recurcive_Karatsuba_ for_1clock* | 6.28 | 1 | 19.35 | 326.15 | 6311.0 |
| *recurcive_Bailey_ for_1clock* | 7.02 | 1 | 16.94 | 441.75 | 7483.3 |

## 5. Conclusions and Outlook

In this paper we discussed the iterative application of Karatsuba's method for polynomial multiplications as a means to reduce the chip area and energy needed to run elliptic curve cryptography on mobile devices. In order to evaluate our approach we analyzed different methods for polynomial multiplication in $GF(2^n)$, and implemented different polynomial multiplication algorithms. For our own approach we realized several partial multipliers. Weused them to implement a set of iterative polynomial multipliers with the goal to identify the one which is best suited for application in mobile devices. Our results clearly indicate that our iterative approach leads to significantly better results with respect to area and energy consumption than the original straight forward application.

Our next step is the finalization of our Montgomery ‚kP' multiplier. In this multiplier we will use the Fermat theorem, since it allows to determine the inverse for the division multiplication and squaring. The Fermat theorem is slower than the Extended Euklidian Algorithm or the method proposed by Shantz [7], but it requires less area. Since the Montgomery method requires only a single division, we think that the smaller area outweighs the slower performance.

## Acknowledgements

This work was partially funded by the German Ministry of Education and Research under grant 01AK060B.